\documentclass[pre]{revtex4}
\usepackage[latin1]{inputenc}
\usepackage{graphics,epsfig,subfigure,changebar,picinpar,rotating}
\usepackage{amsmath,amssymb,amsfonts,amsthm}
\usepackage{tabularx,graphicx,latexsym,bm}
\newlength{\GraphicsWidth}
\renewcommand{\r}{\mathbf{r}}
\newcommand{\tildea}{\tilde{a}}

\newcommand{\Dd}[2]{\frac{\partial{#1}}{\partial{#2}}}

\newcommand{\der}[1]{\frac{d}{d{#1}}}
\newcommand{\derd}[2]{\frac{d{#1}}{d{#2}}}
\newcommand{\derdd}[2]{\frac{d^2{#1}}{d{#2}^2}}

\begin{document}
\title{Screening of charged spheroidal colloidal particles}
\author{Carlos \'Alvarez} 
\email{carl-alv@uniandes.edu.co}
\affiliation{Departamento de F\'{\i}sica, Universidad de los Andes,
  A.A. 4976, Bogot\'a, Colombia}
\affiliation{Laboratoire de Physique Th\'eorique et Mod\`eles
  Statistiques, Universit\'e de Paris-Sud, UMR CNRS 8626, Bâtiment
  100, 91405 Orsay Cedex, France}
\affiliation{Laboratoire de Physique Th\'eorique, 
Universit\'e de Paris-Sud, UMR CNRS 8627, Bâtiment
  210, 91405 Orsay Cedex, France}

\author{Gabriel T\'ellez}
\email{gtellez@uniandes.edu.co} 
\affiliation{Departamento de
  F\'{\i}sica, Universidad de los Andes, A.A. 4976, Bogot\'a,
  Colombia}
\affiliation{Laboratoire de Physique Th\'eorique et Mod\`eles
  Statistiques, Universit\'e de Paris-Sud, UMR CNRS 8626, Bâtiment
  100, 91405 Orsay Cedex, France}


\begin{abstract}
We study the effective screened electrostatic potential created by a
spheroidal colloidal particle immersed in an electrolyte, within the
mean field approximation, using Poisson--Botzmann equation in its
linear and nonlinear forms, and also beyond the mean field by means of
Monte Carlo computer simulation. The anisotropic shape of the particle
has a strong effect on the screened potential, even at large distances
(compared to the Debye length) from it. To quantify this anisotropy
effect, we focus our study on the dependence of the potential on the
position of the observation point with respect with the orientation of
the spheroidal particle. For several different boundary conditions
(constant potential, or constant surface charge) we find that, at
large distance, the potential is higher in the direction of the large
axis of the spheroidal particle.
\end{abstract}

\maketitle

\section{Introduction}
Charged colloidal systems are widespread in nature and their study is
important as they include systems such as clays and protein solutions
in water.  Derjaguin, Landau, Verwey and Overbeek studied the
stability of lyophobic colloids by solving the Poisson--Boltzmann
equation \cite{Verwey,Derjaguin} and stated the theory known as
DLVO. They found the screened electrostatic potential for a spherical
colloidal particle of radius $R$ and valence $Z$, immersed in an
electrolyte with positive ions of charge $z_{+} e$ and density $n_+$,
and negative ions of charge $-z_{-}e$ and density $n_-$. In the
linearized (Debye-H\"uckel) regime, the screened potential is given by
\begin{equation}
\Psi(r)=Zl_B\frac{e^{\kappa R}}{(1+\kappa R)}\frac{e^{-\kappa r}}{r},
\label{yukawa}
\end{equation}
where $\Psi$ denotes the dimensionless potential $\beta
e\times$potential, $l_B=\beta e^2/\epsilon$ is the Bjerrum length, $e$
is the elementary charge, $\beta=(k_BT)^{-1}$, with $T$ the
temperature, $k_B$ the Boltzmann constant, and $\kappa^2=4\pi
l_B(z_+^2 n_++z_{-}^2n_-)$ is the inverse Debye length.

However, many colloidal systems of importance have, in their disperse
phase, macromolecules whose shape cannot be considered spherical, for
example, mineral liquid crystals \cite{Davidson}, and thus their
behavior is not properly explained by the solution of
Poisson--Boltzmann equation for the case of a spherical coarse grained
particle. Therefore it is worthwhile to study models for non spherical
particles, for example ellipsoids of revolution (also known as
spheroids). Furthermore, there are now synthetic clays with uniform
particle shapes like laponite \cite{Dijkstra}, which can be modeled as
oblate spheroids and it is also possible to synthesize spheroidal
nano-particles \cite{Ding}, which provides experimental data for these
models. 

The solution of the linear Poisson--Boltzmann equation has been carried
out for particles with shapes other than spherical (or the infinite
plane case), in particular the cylinder \cite{Tellez,Trizac}, disc
\cite{Agra,Leote,Trizac2} and also the more general case of ellipsoids
of revolution \cite{Hsu,Hsu2} have been studied. For particles with a
disc shape, the screened potential has a long range behavior which
decays with the distance as the Yukawa potential, but multiplied by an
anisotropy function~\cite{Agra}, which characterizes the angular
dependence of the electrostatic potential at large distance. This is
actually a general feature for anisotropic charged objects immersed in
an electrolyte: the screened electrostatic potential is anisotropic
even at large distances from the charged
object~\cite{Chapot-Bocquet-Trizac}. This can be contrasted to the
situation of a charged object in the vacuum, where the electrostatic
potential created by this object becomes isotropic at large distances:
the first term in the multipolar expansion is isotropic. In this work,
we study further this anisotropy function for the general case of
charged ellipsoids of revolution in the Debye-H\"uckel as well as in the
nonlinear Poisson--Boltzmann regimes.

\begin{figure}[ht]
\centering
\epsfig{file=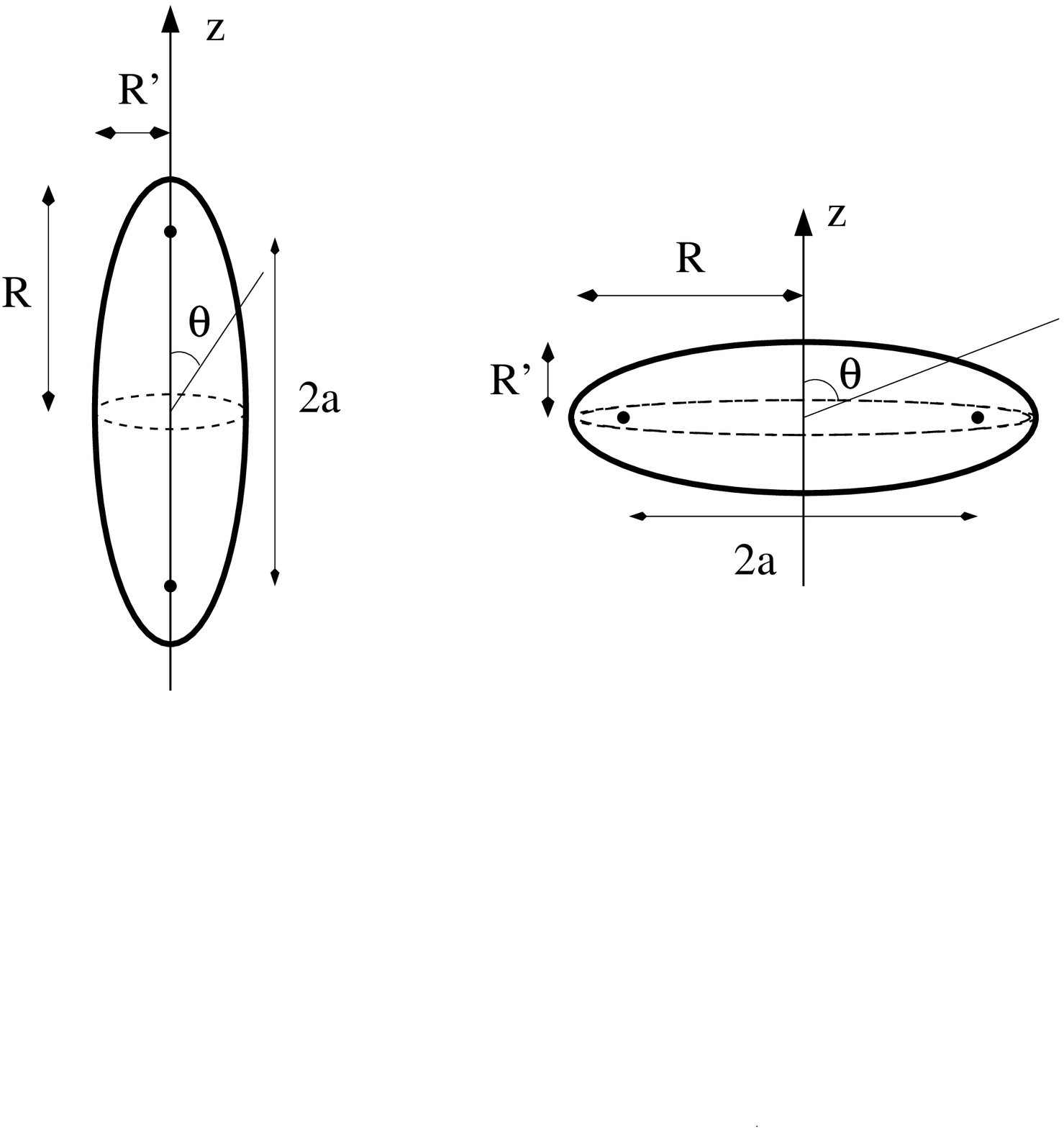,height=10cm}
\vspace{-3cm}
\caption{Prolate (left) and oblate (right) spheroids.}
\label{fig:spheroids}
\end{figure}

For the present geometry of the system the spheroidal coordinates will
be used.  The spheroidal coordinates $(\xi, \eta, \phi)$ are related
to the cartesian coordinates $(x,y,z)$ by \cite{Morse}:
\begin{eqnarray}
x&=&a\sqrt{(\xi^2-1)(1-\eta^2)}\cos\phi\nonumber\\
y&=&a\sqrt{(\xi^2-1)(1-\eta^2)}\sin\phi\nonumber\\
z&=&a\ \xi\ \eta
\label{Pca}
\end{eqnarray}
for the prolate case, and
\begin{eqnarray}
x&=&a\sqrt{(\xi^2+1)(1-\eta^2)}\cos\phi\nonumber\\
y&=&a\sqrt{(\xi^2+1)(1-\eta^2)}\sin\phi\nonumber\\
z&=&a\ \xi\ \eta
\end{eqnarray}
for the oblate case. Here $2a$ is the distance between the foci of the
ellipse. The coordinate $\xi$ is called the radial coordinate since it
plays a similar role as $r$ in spherical coordinates
$(r,\theta,\phi)$. The coordinate $\eta$ will be called angular
coordinate, as it plays a similar role to $\cos\theta$ in the spherical
case. The surface of constant $\xi$ is a spheroid, with
major and minor semi-axes, $R$ and $R'$ given by  
\begin{eqnarray}
R&=&a\ \xi,\nonumber\\
R'&=&a\sqrt{\xi^2-1}
\end{eqnarray}
for the prolate case, and
\begin{eqnarray}
R'&=&a\ \xi,\nonumber\\
R&=&a\sqrt{\xi^2+1}
\end{eqnarray}
for the oblate case, see figure~\ref{fig:spheroids}. Note that the
change of variables $\xi\rightarrow i\xi$ and $a\rightarrow-ia$ in
equations (\ref{Pca}), transforms from the prolate to the oblate
coordinates, so in the following we will treat in detail the prolate
case, and only report the results for the oblate case, without
repeating the calculations.  In section \ref{Lin}, we study the
solution to Poisson--Boltzmann equation in the linearized regime, for
spheroidal colloidal particles in two particular solvable cases:
constant surface charge for an ion-penetrable colloid and constant
surface potential. Section~\ref{nonLin} deals with the numerical
solution of the nonlinear Poisson--Boltzmann equation for the
spheroidal particles. In section \ref{Sim} some Monte Carlo simulation results
for systems of spheroidal colloids with a point like charge at their centers
and explicit ions are presented.

\section{Linear Poisson--Boltzmann equation}\label{Lin}

We consider a single colloidal particle with spheroidal shape,
immersed in the electrolyte. We choose the center of the coordinate
system in the center of the center of the colloidal particle. In
spheroidal coordinates, the surface of the colloidal particle is given
by $\xi=\xi_0$.

In the Debye-H\"uckel regime, Poisson--Boltzmann equation for the
potential becomes
\begin{equation}
\nabla^2\Psi(\xi,\eta,\phi)=\kappa^2 \Psi(\xi,\eta,\phi)
\label{helmh}
\end{equation}
which is separable in spheroidal coordinates. Writing
$\Psi(\xi,\eta,\phi)=A(\xi)B(\eta)C(\phi)$, the equations in each of
the coordinates for the prolate case are
\begin{eqnarray}
\label{eq:sphero1}
\der{\xi}\left[(\xi^2-1)\derd{A}{\xi}\right]-\left(\lambda-\tilde{a}^2\xi^2+\frac{m^2}{\xi^2-1}\right)A=0\\
\label{eq:sphero2}
\der{\eta}\left[(1-\eta^2)\derd{B}{\eta}\right]+\left(\lambda-\tilde{a}^2\eta^2-\frac{m^2}{1-\eta^2}\right)B=0\\
\derdd{C}{\phi}+m^2C=0,\label{phi}
\end{eqnarray}
with $\tilde{a}=\kappa a$. The solutions for (\ref{phi}) are
\begin{equation}
C(\phi)=e^{\pm im\phi}.
\end{equation}
The functions of the $\xi$ and $\eta$ coordinates (equations
(\ref{eq:sphero1}) and (\ref{eq:sphero2})) satisfy the same differential equation, known as
the spheroidal equation, which is a generalization of Legendre
equation. The only difference between the radial and angular functions
is the range in which they are defined ($-1\leq\eta\leq 1$, and
$1\leq\xi<\infty$ in the prolate case, or $0\leq\xi<\infty$ in the
oblate case). The constant of integration $\lambda$ is a function of
integers $l$ and $m$, and in the spherical (Legendre) case it is
$\lambda=l(l+1)$, for $l=0,1,2, \ldots$. The solutions (radial and
angular) are expressed as expansions in the Legendre functions and are
known as the spheroidal wave functions. We use the notation and
normalization presented in~\cite{Meixner, Falloon} for the spheroidal
wave functions.  In this work, the numerical calculation of the
spheroidal wave functions is done with {\it Mathematica}, which uses
the same conventions as in~\cite{Meixner, Falloon}. The angular
functions are denoted $ps_l^m$. They are the analogous to the
associate Legendre functions $P_{l}^m$ in the spherical case. The
normalization used here is~\cite{Meixner,Falloon}
\begin{equation}
\int_{-1}^1|ps_l^m(i\tilde{a},\eta)|^2d\eta=\frac{2}{2l+1}\frac{(l+m)!}{(l-m)!}.
\end{equation}
There are two groups of solutions for the radial equation
$S_l^{m(1)}$ and $S_l^{m(2)}$. For $\tilde{a}\xi\rightarrow\infty$
these functions behave as the spherical Bessel functions
\begin{eqnarray}
S_l^{m(1)}(i\tilde{a},\xi)&\sim& j_l(i\tilde{a}\xi)\sim
\frac{1}{i\tilde{a}\xi}\cos(i\tilde{a}\xi-l\pi/2),\label{s1}\\
S_l^{m(2)}(i\tilde{a},\xi)&\sim& n_l(i\tilde{a}\xi)\sim
\frac{1}{i\tilde{a}\xi}\sin(i\tilde{a}\xi-l\pi/2).\label{s2}
\end{eqnarray}
To satisfy the boundary condition $\Psi(\xi,\eta,\phi)\to0$ when
$\xi\to\infty$, the appropriate radial function $R_l^m$ is the
combination
\begin{equation}
R_l^m(i\tilde{a},\xi)=-i^l\left[S_l^{m(1)}(i\tilde{a},\xi)+iS_l^{m(2)}(i\tilde{a},\xi)\right],
\label{rad}
\end{equation}
which is analogous to the spherical Hankel function for the spherical
case. The solution to equation (\ref{helmh}) is now expressed as a
superposition of spheroidal wave functions
\begin{equation}
\Psi(\xi,\eta,\phi)=\sum_{l,m}^{\infty}a_l^me^{im\phi}ps_l^m(i\tilde{a},\eta)R_l^m(i\tilde{a},\xi).
\label{potP}
\end{equation}
for the prolate case. In the oblate situation,
\begin{equation}
\Psi(\xi,\eta,\phi)=\sum_{l,m}^{\infty}a_l^me^{im\phi}ps_l^m(\tilde{a},\eta)
R_l^m(\tilde{a},i\xi).
\label{eq:potPoblat}
\end{equation}

From equations (\ref{s1}) and (\ref{s2}), it can be seen that the
behavior of the radial functions of equation (\ref{rad}) as
$\xi\rightarrow\infty$ is the same as the one of the screened
potential (eq.~(\ref{yukawa})). Indeed, when $\xi\to\infty$, we have
$a\xi\sim r$, and $\eta\sim\cos\theta$, and
\begin{equation}
R_l^m(i\tilde{a},\xi)\sim\frac{e^{-\kappa r}}{\kappa r}\,.
\end{equation}
Therefore, far from the colloid surface, $\tildea\xi\gg1$, the potential
behaves as
\begin{equation}
\label{eq:potential-anisotrop}
\Psi(r,\theta,\phi)\sim
Z\ l_B\ f(\tilde{a},\theta,\phi)\frac{e^{-\kappa r}}{r},
\end{equation}
where the anisotropy function $f(\tilde{a},\theta,\phi)$ is
given by
\begin{equation}
  \label{eq:anisotropy-gen}
  f(\tilde{a},\theta)=
  \frac{1}{Z\ l_B\kappa}\sum_{l,m}a_l^me^{im\phi}ps_l^m(i\tilde{a},\cos\theta).
\end{equation} 
In this work, we will only consider spheroids with a surface charge
density which is rotationally invariant around the $z$ axis. Therefore
the potential will be independent of the azimuthal coordinate $\phi$
and only the terms with $m=0$ survive in
equation~(\ref{eq:anisotropy-gen}).

To characterize the anisotropy along the minor and major axis
directions, we define a ``maximum anisotropy function''
\begin{equation}
\label{eq:max-anisotrop}
f_M(\tilde{a})=\frac{\left|f(\tilde{a},\pi/2)-f(\tilde{a},0)\right|}{\min[f(\tilde{a},\pi/2),f(\tilde{a},0)]}.
\end{equation} 
In the next section, we will compute the anisotropy function for
various aspect ratios maintaining the size of the major semi-axis
($R$) constant.

\subsection{Constant surface charge density}

A case which allows for an analytical solution to Poisson--Boltzmann
equation in the linear Debye-H\"uckel regime is when the colloidal
particle is penetrable by the ions of the electrolyte \cite{Hsu} and
it has a constant charge density $\sigma$ in its surface
$\xi=\xi_0$. This model can be applied to cells or vacuoles with ion
channels, or as a coarse grained description of a globule formed by a
hydrophobic polyelectrolyte~\cite{Trizac-hydrophobic-poly} in a poor
solvent. The total charge of the colloid is
\begin{equation}
Ze=\sigma
\int dS=2\pi\sigma\,a^2\sqrt{\xi_0^2-1}
\int_{-1}^1\sqrt{\xi_0^2-\eta^2}d\eta=
2\pi\sigma\,a^2\sqrt{\xi_0^2-1}
\left(\sqrt{\xi_0^2-1}+\xi^2\tan^{-1}\frac{1}{\sqrt{\xi_0^2-1}}
\right)
\,.
\label{ZlconstQ}
\end{equation}

In this case, the electrostatic potential satisfies the linear
Poisson--Boltzmann equation~(\ref{helmh}), inside and outside the
spheroid, with the same Debye length $\kappa^{-1}$. Therefore, the
potential inside of the particle has the following form, in the
prolate case,
\begin{equation}
\Psi_{in}(\xi,\eta)=
\sum_{l}^{\infty}b_l^0ps_l^0(i\tilde{a},\eta)S_l^{0(1)}(i\tilde{a},\xi).
\end{equation}
Here, only the radial function $S_l^{m(1)}$ is used for the radial
part, since $S_l^{m(2)}$ diverges as $\xi$ goes to $1$ (prolate case) or
$0$ (oblate case). On the other hand, the potential outside of the
particle has the form shown in equation (\ref{potP}), with
$m=0$.

The coefficients $a_{l}^{0}$ and $b_{l}^{0}$ are obtained by imposing
the boundary conditions that the electrostatic potential at the
surface of the spheroid $\xi=\xi_0$ is continuous, and that the
discontinuity of the normal ($\mathbf{n}$) component of the electric
field at the spheroid surface is proportional to its surface charge
density
\begin{eqnarray}
  4\pi l_B \sigma/e &=& \nabla \Psi(\xi_0^{-},\eta) \cdot \mathbf{n} - \nabla
  \Psi(\xi_0^{+},\eta) \cdot \mathbf{n} \nonumber\\
  &=&
  \frac{1}{a}\frac{\sqrt{\xi_0^2-1}}{\sqrt{\xi_0^2-\eta^2}}
  \left[
    \frac{\partial\psi}{\partial\xi}(\xi_0^{+},\eta)
    - \frac{\partial\psi}{\partial\xi}(\xi_0^{-},\eta)
    \right]
  \label{eq:discont-psi-prime}
\end{eqnarray}

From the continuity of the electrostatic potential, we obtain
\begin{equation}
b_l^0=\frac{R_l^0(i\tilde{a},\xi_0)}{S_l^{0(1)}(i\tilde{a},\xi_0)}a_l^0.
\end{equation}
Using equation~(\ref{eq:discont-psi-prime}) and the orthogonality of
the angular spheroidal wave functions
\begin{equation}
\int_{-1}^1 ps_l^m(i\tilde{a},\eta)ps_{l'}^m(i\tilde{a},\eta)d\eta=\frac{2}{2l+1}\frac{(l+m)!}{(l-m)!}\,\delta_{ll'}
\end{equation}
we obtain the coefficients $a_{l}^0$ 
\begin{eqnarray}
a_l^0=2\pi\frac{l_B\ \sigma}{ei^l}\,\kappa a^2\,
S_l^{0(1)}(i\tilde{a},\xi_0)\sqrt{\xi_0^2-1}\,(2l+1)
\int_{-1}^1\sqrt{\xi_0^2-\eta^2}\,ps_l^0(i\tilde{a},\eta)d\eta\,.
\label{aconstQ}
\end{eqnarray}
The angular spheroidal wave functions $ps_{l}^{0}$ have the same
parity as $l$, therefore for $l$ odd, the coefficient $a_{l}^{0}=0$.
Replacing (\ref{aconstQ}) and (\ref{ZlconstQ})
into~(\ref{eq:anisotropy-gen}), we obtain the anisotropy function for
the prolate case
\begin{eqnarray}
\label{eq:faniso-prolat-sigma-cst}
f(\tilde{a},\theta)_{pro}=
\frac{
\sum_l(2l+1)i^{-l}S_l^{0(1)}(i\tilde{a},\xi_0)ps_l^0(i\tilde{a},\cos\theta)
\int_{-1}^1\sqrt{\xi_0^2-\eta^2}ps_l^0(i\tilde{a},\eta)d\eta}{
\sqrt{\xi_0^2-1}+\xi_0^2\tan^{-1}(1/\sqrt{\xi_0^2-1})}\,.
\end{eqnarray}
In the oblate case, it is
\begin{eqnarray}
\label{eq:faniso-oblate-sigma-cst}
f(\tilde{a},\theta)_{obl}=\frac{
\sum_l(2l+1)i^{-l}S_l^{0(1)}(\tilde{a},i\xi_0)ps_l^0(\tilde{a},\cos\theta)
\int_{-1}^1\sqrt{\xi_0^2+\eta^2}ps_l^0(\tilde{a},\eta)d\eta}{\sqrt{\xi_0^2+1}+\xi_0^2\ln\frac{1+\sqrt{1+\xi_0^2}}{\xi_0}}
\,.
\end{eqnarray}
The sum over $l$ runs through all even integers.

\begin{figure}[ht]
\centering
\epsfig{file=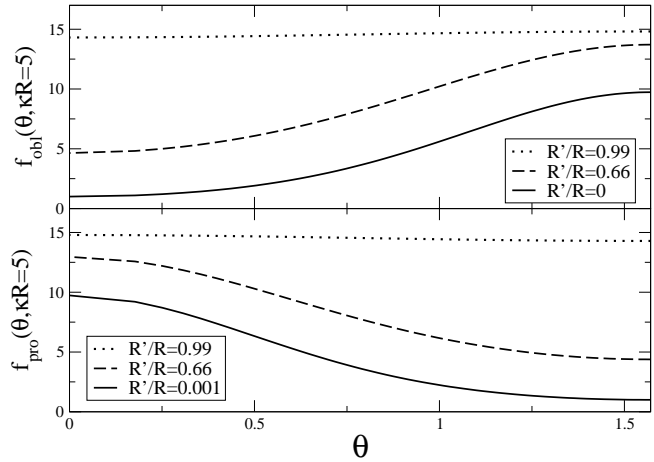,height=10cm,angle=-90}
\caption{Anisotropy function for the oblate(top) and prolate(bottom)
  cases for various aspect ratios ($R'/R$) and a constant surface
  charge density. Here the major semi-axis $R$ is five times the Debye
  length.}
\label{anisQ}
\end{figure}

Figure~\ref{anisQ} shows the anisotropy function for the constant
surface charge case as the aspect ratio ($R'/R$) changes from the
sphere to the rod for the prolate case and from the sphere to the disc
for the oblate case.

For a prolate spheroid, the anisotropy function, and thus the
electrostatic potential at large distances, is larger in the direction
$\theta=0$ of the large axis. For an oblate spheroid, the anisotropy
function is also larger in the direction of the large axis (now
labeled $\theta=\pi/2$).

\begin{figure}[ht]
\centering
\epsfig{file=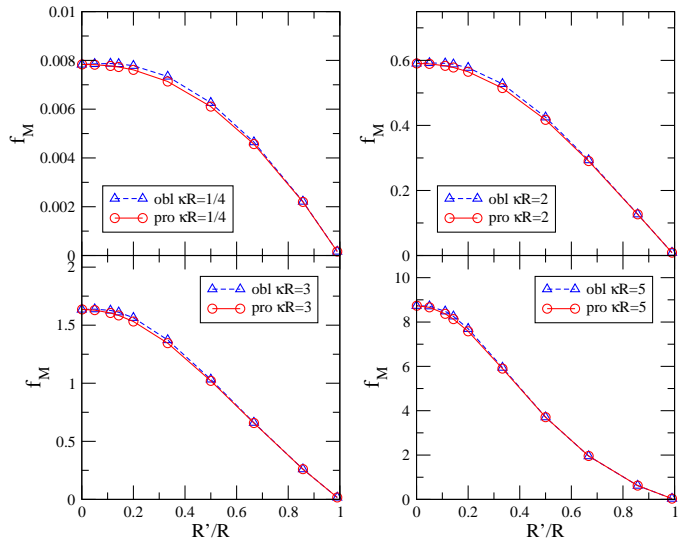,height=12cm,angle=-90}
\caption{Maximum anisotropy function for the oblate and prolate cases for
  various aspect ratios and Debye length values with a constant surface charge
  density.}
\label{ManisQ}
\end{figure}

Now, we study the difference in the potential along the major and
minor axes directions ($\theta=0$ and $\theta=\pi/2$), with the
maximum anisotropy function defined in
Eq.~(\ref{eq:max-anisotrop}). The maximum anisotropy function for
various values of the Debye length and different aspect ratios is
shown on figure~\ref{ManisQ}. It can be seen that the function is
small when the Debye length is greater than the dimensions of the
particle, as the big cloud of ions screens the charge on the colloid
surface in such a way that far from the colloid the potential is
almost isotropic. On the other hand, for small values of the Debye
length the potential retains its anisotropy even at large
distances. It can also be seen that the shape of $f_M$ vs.~aspect
ratio changes its curvature as the Debye length is increased.

Another boundary condition of practical interest is when the colloidal
particle has constant surface charge density but it is impenetrable by
the electrolyte. Unfortunately, this situation turns out to be more
difficult to treat analytically. The reason is the following. Inside
the spheroidal particle, the electrostatic potential satisfies Laplace
equation. In spheroidal coordinates, the corresponding solution to
Laplace equation is~\cite{Morse}
\begin{equation}
  \label{eq:Laplacesol}
  \Psi_{\text{in}}(\xi,\eta)=\sum_{l} d_l P_{l}(\xi) P_{l}(\eta)
\end{equation}
where $P_{l}$ are the Legendre polynomials. The potential
inside~(\ref{eq:Laplacesol}) and outside~(\ref{potP}) are expanded in
different basis of functions for the angular coordinate $\eta$. To
satisfy the boundary conditions at the surface $\xi=\xi_0$ of the
spheroid and find the coefficients $d_{l}$ and $a_{l}^{0}$, one has
first to change the basis of angular functions to use the same basis
inside and outside, i.e. express the Legendre polynomials in terms of
the spheroidal wave functions $ps_{l}^{0}$ or vice-versa. We will not
attempt to follow this route. For most practical situations, the
colloidal particle will be highly charged. In that situation, the
nonlinear Poisson--Boltzmann equation should be used. As we will
explain in the following section and in section~\ref{nonLin}, for
highly charged colloids, the linear Poisson--Boltzmann equation with
Dirichlet boundary conditions (fixed constant potential at the surface
of the particle) is more relevant.

\subsection{Constant surface potential}
\label{subsec:constant-pot}

Now, we turn our attention to the case when the colloidal particle is
maintained at a fixed constant potential and it is not penetrable by
the electrolyte. This situation is relevant for the case of highly
charged colloidal particles. Indeed, as explained
in~\cite{Trizac-renorm-PRL, Trizac-renorm-JCP, Agra}, for highly
charged colloidal particles, the constant potential boundary condition
can be seen as an effective boundary condition to join the solution of
the linear Poisson--Boltzmann equation valid far from the colloid,
with the solution of the nonlinear Poisson--Boltzmann equation valid
near the colloid surface. As a first approximation, the value $\Psi_0$
for the surface potential is obtained using the known solution of
nonlinear Poisson--Boltzmann equation for a planar highly charged
surface. For a charge-symmetric electrolyte it is $\Psi_0=4$. 

Outside the colloidal particle, the potential is given, for a prolate
spheroid, by equation~(\ref{potP}), or, for an oblate spheroid, by
equation~(\ref{eq:potPoblat}). Imposing $\Psi(\xi_0,\eta)=\Psi_0$, and
using the orthogonality of the angular spheroidal wave functions, we
obtain the coefficients
\begin{equation}
a_l^0=\Psi_0\frac{2l+1}{2R_l^0(i\tilde{a},\xi_0)}\int_{-1}^1ps_l^0(i\tilde{a},\eta)d\eta,
\end{equation}
in the prolate case. Due to the parity properties of $ps_{l}^{0}$, if
$l$ is odd, then $a_{l}^{0}=0$.

The anisotropic surface charge density of the colloidal particle is
\begin{eqnarray}
   \sigma(\eta) &=& -\frac{e}{4\pi l_B
     a}\frac{\sqrt{\xi_0^2-1}}{\sqrt{\xi_0^2-\eta^2}}
   \frac{\partial\psi}{\partial\xi}(\xi_0^{+},\eta) \nonumber\\ &=&
   -\frac{e}{4\pi l_B a}\frac{\sqrt{\xi_0^2-1}}{\sqrt{\xi_0^2-\eta^2}}
   \sum_{l} a_{l}^{0} {R_{l}^{0}}'(i\tildea,\xi_0)
   ps_{l}^{0}(i\tildea,\eta)\,,
\label{cpCdens}
\end{eqnarray}
where the prime denotes the derivative with respect to
$\xi$. Integrating over the surface of the colloidal particle, we
obtain its total charge
\begin{equation}
Z\ l_B=-
\frac{a(\xi_0^2-1)}{2}
\sum_l a_l^0{R_l^0}\bm{'}(i\tilde{a},\xi_0)
\int_{-1}^1ps_l^0(i\tilde{a},\eta)d\eta,
\end{equation}
Let us define
\begin{equation}
C_{l}(z)=\int_{-1}^1ps_l^0(z,\eta)d\eta\,,
\label{eq:Cl}
\end{equation}
which is nonzero only when $l$ is even. Then, the charge of the
colloid is
\begin{equation}
Z\ l_B=-
\frac{a\Psi_0(\xi_0^2-1)}{4}
\sum_l (2l+1) C_{l}(i\tildea)^2
\frac{{R_l^0}{'}(i\tilde{a},\xi_0)}{R_{l}^{0}(i\tildea,\xi_0)}
\end{equation}
Finally, the anisotropy function, for the prolate situation is
\begin{eqnarray}
  \label{eq:fanisotrop-prolat-fixed-pot}
f(\tilde{a},\theta)_{pro}
=-
\frac{2}{\tilde{a}(\xi_0^2-1)}
\frac{\displaystyle\sum_l\frac{2l+1}{R_l^0(i\tilde{a},\xi_0)}
C_{l}(i\tildea)\,ps_l^0(i\tilde{a},\cos\theta)}{\displaystyle
  \sum_l\frac{(2l+1)C_{l}(i\tildea)^2\,{R_l^0}\bm{'}(i\tilde{a},\xi_0)}{
    R_l^0(i\tilde{a},\xi_0)}}\,,
\end{eqnarray}
and, for the oblate case, it is
\begin{eqnarray}
  \label{eq:fanisotrop-oblat-fixed-pot}
f(\tilde{a},\theta)_{obl}=-\frac{2}{\tilde{a}(\xi_0^2+1)}
\frac{\displaystyle
\sum_l\frac{2l+1}{R_l^0(\tilde{a},i\xi_0)}
C_{l}(\tildea)\,ps_l^0(\tilde{a},\cos\theta)}
{\displaystyle
\sum_l\frac{(2l+1) C_{l}(\tilde{a})^2\,{R_l^0}\bm{'}(\tilde{a},i\xi_0)}{
R_l^0(\tilde{a},i\xi_0)}}.
\end{eqnarray}

\begin{figure}[ht]
\centering
\epsfig{file=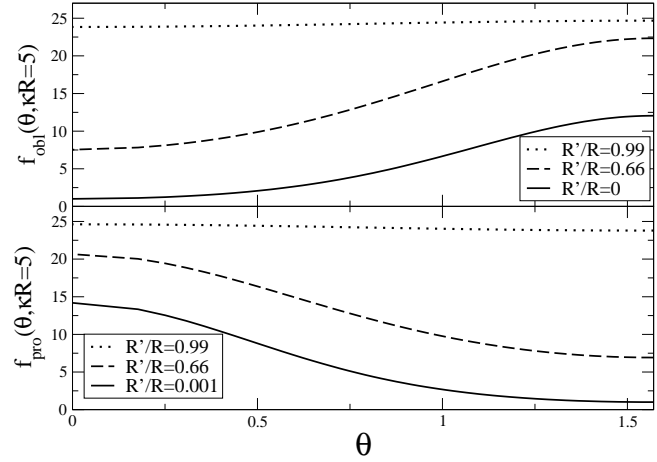,height=10cm,angle=-90}
\caption{Anisotropy function for the oblate(top) and prolate(bottom)
  cases for various aspect ratios ($R'/R$) and a constant surface
  potential. Here the major semi-axis is five times the Debye length.}
\label{anis}
\end{figure}

Figure~\ref{anis} shows the anisotropy function for different aspect
ratios $R'/R$ for the constant surface potential case. Again, the
anisotropy function is larger in the direction of the major axis of
the spheroid, as it was in the case of the constant surface charge
density. However, the overall magnitude of the anisotropy function is
larger in the present case of constant surface potential than in the
constant surface charge situation of the previous section. This can be
understood as a ``tip'' effect. Indeed, with a constant surface
potential boundary condition, the tips of the spheroids acquire a
larger surface charge density, which translate into a higher
anisotropy function in that direction (the large axis direction).

\begin{figure}[ht]
\centering
\epsfig{file=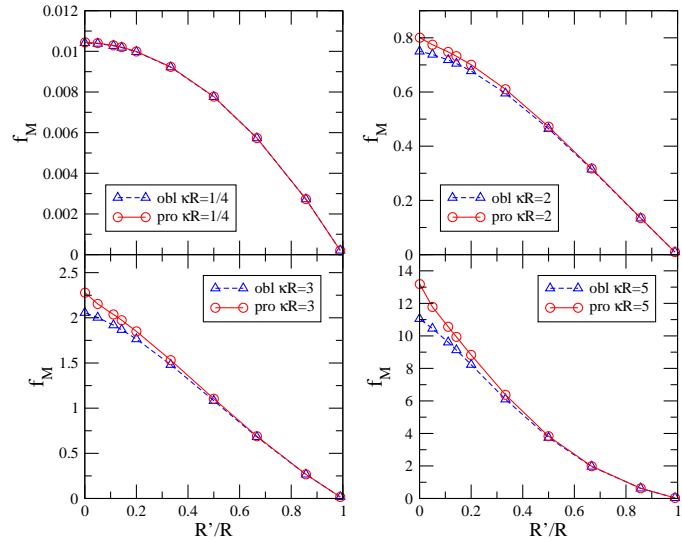,height=12cm,angle=-90}
\caption{Maximum anisotropy function for the oblate and prolate cases for
  various aspect ratios and Debye length values with a constant potential at
  the surface.}
\label{Manis}
\end{figure}

Figure~\ref{Manis} shows the maximum anisotropy function for various
values of the Debye length and different aspect ratios. Comparing it
to figure~\ref{ManisQ} it can be seen that for the present case when
the potential at the surface of the colloid is constant the values of
the maximum anisotropy function for the prolate and oblate cases
become different for small Debye lengths (compared to the large
semi-axis) and for highly anisotropic objects, i.e. small $R'/R$
values, that is, objects near the limiting cases of the disc or the
rod.

\begin{figure}[ht]
\centering
\epsfig{file=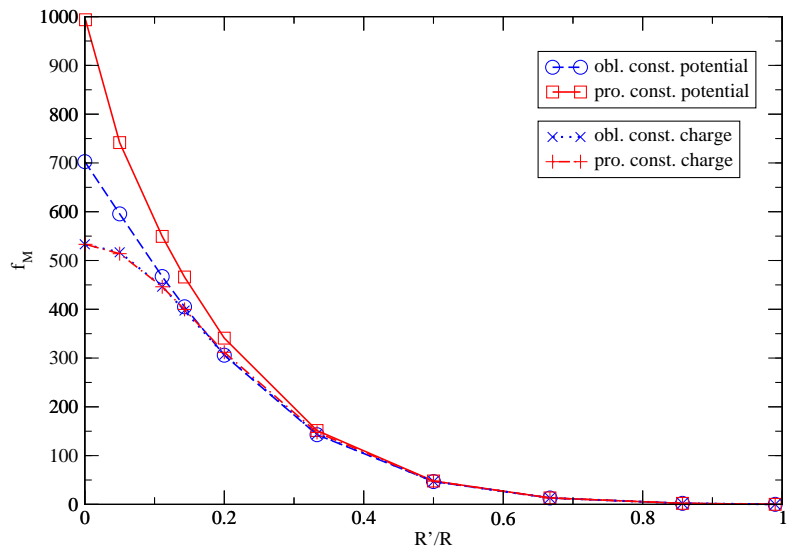,height=12cm,angle=-90}
\caption{Maximum anisotropy function for the oblate and prolate cases for
  $\kappa R=10$.}
\label{anis10}
\end{figure}

Figure~\ref{anis10} compares the maximum anisotropy for the oblate and
prolate cases for a large value of $\kappa R$, and for the two
different situations of constant surface charge density or constant
surface potential. It is clear that at constant surface charge density
the function $f_M$ behaves similarly for the prolate and oblate cases
while at constant surface potential the anisotropy is greater in the
prolate case. This is probably related again to a ``tip'' effect, as
mentioned earlier. In the Dirichlet boundary condition case, the
surface charge at the tips will be more important for prolate
spheroids than for oblate spheroids, since the local curvature at the
tip is higher in the prolate case. This translate into a higher
anisotropy function in the large axis direction for prolate spheroids,
as shown in figure~\ref{anis10}. On the other hand, the constant
surface charge density situation does not discriminate between prolate
and oblate spheroids as far as the anisotropy function is concerned.

\subsection{Special cases: the disc and the rod}

In this section, we give some additional details on two special cases,
the disc and the rod geometries. A disc can be seen as an oblate spheroid
with its surface given by $\xi_0=0$. Disc shaped colloidal particles
have been studied in~\cite{Trizac-Hansen, Leote, Rowan-Hansen-Trizac,
  Trizac2, Agra} as they are an appropriate model for clay suspensions,
like laponite. In~\cite{Trizac2}, the screened electrostatic potential
was studied within the linear Poisson--Boltzmann theory, with Neumann
boundary conditions, i.e. constant surface charge density on the
disc. At large distances from the disc, compared to the Debye length,
the potential is of the form~(\ref{eq:potential-anisotrop}): a Yukawa
screened potential, multiplied by an anisotropy function. The
anisotropy function was computed exactly in~\cite{Trizac2}, for this
situation
\begin{equation}
  f(\tildea,\theta)_{\text{disc,Neumann}}=\frac{2I_1(\tildea \sin\theta)}{\tildea
    \sin\theta} 
  \,.
\end{equation}
Here $\tildea=\kappa a$, with $a$ the radius of the disc, and $I_1$
the modified Bessel function of order 1. Comparing to the general
expression~(\ref{eq:faniso-oblate-sigma-cst}) for oblate spheroids,
and taking the limit $\xi_0\to0^{+}$, we obtain an interesting
expansion formula for the modified Bessel function in terms of
spheroidal wave functions
\begin{equation}
  \sum_{l=0, 2, \ldots}^{\infty} (2l+1)\left[\int_{0}^{1} |\eta'|
    ps_{l}^{0}(\tildea,\eta')\,d\eta'\right]
  i^{-l}S_{l}^{0(1)}(i0^{+},\tildea) ps_{l}^{0}(\tildea,\eta) =
  \frac{I_1(\tildea\sqrt{1-\eta^2})}{\tildea \sqrt{1-\eta^2}} \,.
\end{equation}

\begin{figure}[ht]
\centering
\vspace{1cm}
\includegraphics[width=\GraphicsWidth]{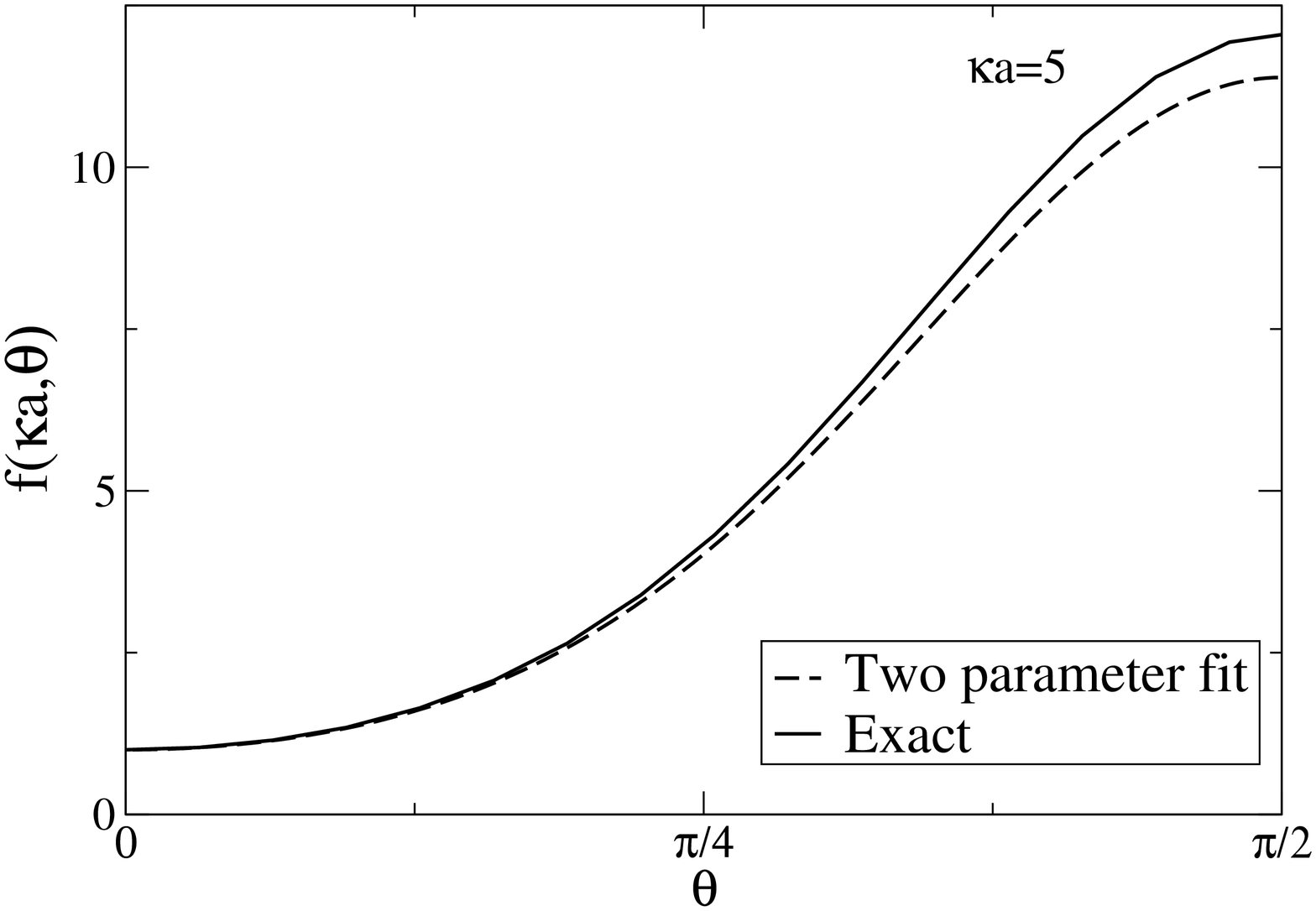}
\caption{Anisotropy function for a disc at constant surface potential
  with radius 5 times the Debye length.}
\label{disc-anis}
\end{figure}

For the case of a disc held at constant surface potential, our work
gives additional information to the one found in the current
literature~\cite{Agra}. In~\cite{Agra}, this problem was studied,
using cylindrical coordinates. Unfortunately, in cylindrical
coordinates, the linear Poisson--Boltzmann equation turns out to be
untractable analytically, when the potential at the surface of the
disc is fixed. The technical difficulty is that the problem involves
mixed boundary conditions when cylindrical coordinates $(\rho,\phi,z)$
are used: a Dirichlet boundary condition for $\rho<a$ and $z=0$, and a
Neumann boundary condition for $\rho>a$ and $z=0$. Using oblate
spheroidal coordinates, this problem disappears, only a Dirichlet
boundary condition is imposed at $\xi=0$. Specializing
equation~(\ref{eq:fanisotrop-oblat-fixed-pot}) for the disc case
($\xi_0\to0^{+}$), we obtain an analytic expression for the anisotropy
function for the disc shaped colloidal particles at fixed potential
\begin{eqnarray}
  \label{eq:fanisotrop-disc-fixed-pot}
  f(\tilde{a},\theta)_{\text{disc,Dirichlet}}=-\frac{2}{\tilde{a}}
  \frac{\displaystyle \sum_l\frac{2l+1}{R_l^0(\tilde{a},i0^{+})}
    C_{l}(\tildea)\,ps_l^0(\tilde{a},\cos\theta)} {\displaystyle
    \sum_l\frac{(2l+1)
      C_{l}^2(\tildea)\,{\partial_{\xi}R_l^0}(\tilde{a},i0^{+})}{
      R_l^0(\tilde{a},i0^{+})}}
  \,.
\end{eqnarray}
An analytical expression for the anisotropy function in this situation
was previously unknown. Exactly, it was only known that~\cite{Agra}
$f(\tilde{a},0)_{disc,Dirichlet}=1$. This can be verified numerically
in the previous expression, and gives yet another interesting relation
between the spheroidal wave functions
\begin{equation}
  ps_{l}^{0}(1,\tildea)=-\frac{\tildea}{2}\,
  \int_{-1}^{1} ps_l^0(\tildea,\eta)\,d\eta\,
  \left.\frac{\partial
    R_{l}^{0}}{\partial\xi}(\tildea, i\xi)\right|_{\xi\to{0}}
  \,.
\end{equation}
Since an analytical expression for the anisotropy function in this
case was previously unavailable, the authors of~\cite{Agra} proposed
an approximate analytical ansatz based on a two parameter fit,
equation~(16) of Ref.~\cite{Agra}. In figure~\ref{disc-anis}, we
compare the ansatz from~\cite{Agra}, with the exact analytical
expression~(\ref{eq:fanisotrop-disc-fixed-pot}). As it can be seen in
the figure, the two parameter ansatz proposed in~\cite{Agra}, is a
fairly good approximation. It underestimates the anisotropy, but, for
$\tildea=5$, the relative maximum error is for $\theta=\pi/2$, and it
is only 5\%.

The screening of a rod-like stiff polyelectrolyte, such DNA fragments,
Tobacco Mosa\"ic Virus (TVM), etc..., can be described by our general
approach, since a rod can be seen as a prolate spheroid with
$\xi_0=1$. In~\cite{Chapot-Bocquet-Trizac, Chapot-Bocquet-Trizac2},
the screened electrostatic potential and the anisotropy function of
rod-like polyelectrolytes, modeled as cylinders of finite length and
nonzero radius, were computed numerically. From our approach, we
obtain analytical results for the limiting case of infinitely thin
rods (radius zero). For a thin rod with fixed charge density, the
anisotropy function can be obtained
from~(\ref{eq:faniso-prolat-sigma-cst}), taking the limit $\xi_0\to
1$,
\begin{equation}
  f(\tildea, \theta)_{\text{rod,Neumann}}=\frac{2}{\pi}\,
  \sum_{l}(2l+1)i^{-l} S_l^{0(1)}(1,i\tildea) \left[\int_{-1}^{1}
    \sqrt{1-\eta'^2}ps_{l}^{0}(\tildea,\eta')\,d\eta'\right]
  ps_{l}^{0}(\tildea,\cos\theta)
  \,.
\end{equation}
Here $2a$ is the length of the rod.

The case of a rod at fixed constant potential is somehow pathological,
as the spheroidal wave function $R_{l}^{0}(i\tildea,\xi_0)$ diverges
as $\xi_0\to 1$. This difficulty already happens in the absence of the
electrolyte. A charged rod in the vacuum having a fixed constant
potential would have an infinite linear charge density. Indeed, in the
vacuum, the potential of a prolate spheroid with fixed potential
$\Psi_0$ on its surface ($\xi=\xi_0$) is~\cite{Morse}
\begin{equation}
  \Psi_\text{{prolate,vacuum}}(\xi)=\Psi_0\, \frac{\ln\frac{\xi+1}{\xi-1}
  }{\ln\frac{\xi_0+1}{\xi_0-1}}
  \,.
\end{equation}
This expression has a logarithmic divergence when $\xi_0\to 1$. The
corresponding surface charge density is
\begin{equation}
  \sigma(\eta)_\text{{vacuum}}=
  \frac{\Psi_0}{\pi a \sqrt{(\xi_0^2-\eta^2)(\xi_0^2-1)}
    \,\ln\frac{\xi_0+1}{\xi_0-1}}
\end{equation}
In the limit of a rod, $\xi_0\to1$, one can define the corresponding
linear charge density $\lambda(\eta)$, remembering that the surface
element is $dS=a^2\sqrt{(\xi_0^2-\eta^2)(\xi_0^2-1)}\,d\eta\,d\phi$,
\begin{equation}
  \lambda(\eta)_{\text{vacuum}}=\lim_{\xi_0\to1}\pi a
  \sqrt{(\xi_0^2-\eta^2)(\xi_0^2-1)}\,\sigma(\eta)=\lim_{\xi_0\to1}
  \frac{\Psi_0}{\ln\frac{\xi_0+1}{\xi_0-1}}\,.
\end{equation}
The linear charge density has the same logarithmic divergence when
$\xi\to 0$ as the potential.

This same difficulty appears in the presence of an electrolyte for the
screened potential, the potential and the charge have a logarithmic
divergence when $\xi_0\to1$. Nevertheless, the anisotropy function
will have a finite limit when $\xi_0\to1$, since it is defined as the
ratio between the potential at large distance and the total charge of
the rod: the logarithmic divergences will cancel.  When $\xi_0\to1$,
\begin{equation}
  \label{eq:asymptR}
  R_{l}^{0}(i\tildea,\xi_0)\sim g_{l}(\tildea) \ln(\xi_0-1)\,.
\end{equation}
We shall determine the explicit expression for $g_l(\tildea)$ later
(equation~(\ref{eq:g})). The derivative with respect to $\xi$
diverges as
\begin{equation}
  {R_{l}^{0}}\bm{'}(i\tildea,\xi_0)\sim \frac{g_{l}(\tildea)}{\xi_0-1}
\end{equation}
Therefore, in the limit $\xi_0\to 1$, the anisotropy
function~(\ref{eq:fanisotrop-prolat-fixed-pot}) has a finite limit
\begin{equation}
  f(\tildea,\theta)_\text{{rod,Dirichlet}}=
  \frac{\sum_{l}(2l+1)\,C_{l}(i\tildea)\,
    ps_{l}^{0}(i\tildea,\cos\theta)/g_{l}(\tildea)}{
    \sum_{l}(2l+1)\,C_{l}^2(i\tildea)}
\end{equation}
with $C_{l}(i\tildea)$ given by~(\ref{eq:Cl}). The prefactor
$g_{l}(\tildea)$ in the asymptotic formula~(\ref{eq:asymptR}) can be
determined as follows. In the direction perpendicular to the rod,
$\theta=\pi/2$, the anisotropy function is equal to 1. Therefore
\begin{equation}
  \label{eq:g}
  g_{l}(\tildea)=-\frac{ps_{l}^{0}(i\tildea,0)}{\tildea\,C_l(i\tildea)}
  \,.
\end{equation}
Finally,
\begin{equation}
  f(\tildea,\theta)_\text{{rod,Dirichlet}}=
  \frac{\sum_{l}(2l+1)\,C_{l}(i\tildea)^{2}\,
    \displaystyle
    \frac{ps_{l}^{0}(i\tildea,\cos\theta)}{ps_{l}^{0}(i\tildea,0)}}{
    \sum_{l}(2l+1)\,C_{l}(i\tildea)^2}
  \,.
\end{equation}
As a subproduct of this analysis, we obtained the asymptotic behavior of
the spheroidal wave function, when $\xi\to 1$,
\begin{equation}
  \label{eq:asymptRxi1}
  R_{l}^{0}(i\tildea,\xi)\mathop{\sim}\limits_{\xi\to1}
  -\frac{ps_{l}^{0}(i\tildea,0)}{\tildea\,
    C_{l}(i\tildea)}\,\ln(\xi-1)\,.
\end{equation}

\section{Nonlinear Poisson--Boltzmann equation}\label{nonLin}

\subsection{Renormalized anisotropy function}

When the colloidal particle is highly charged, Poisson--Boltzmann
equation cannot be linearized. The full nonlinear equation reads
\begin{equation}
  \label{PBnl}
  \nabla^2 \Psi =  \frac{\kappa^2}{z_{+}+z_{-}}
  \left[ e^{z_{-} \Psi}- e^{-z_{+} \Psi} \right]
  \,.
\end{equation}
Although near the surface of the highly charged colloidal particle,
the full nonlinear equation should be used, at large distances from
it, the potential becomes small due to the screening. Therefore, at
large distances, compared to the Debye length, from the colloidal
particle, the potential satisfies again the linear
equation~(\ref{helmh}). For a spherical symmetric colloidal particle,
this means that the potential at large distances from the particle
will be of the DLVO form~(\ref{yukawa})
\begin{equation}
  \Psi(r)\sim Z_{\text{ren}} l_B \frac{e^{\kappa R}}{(1+\kappa
    R)}\,\frac{e^{-\kappa r}}{r}
  \,,
\end{equation}
but now, $Z_{\text{ren}}$ is not the bare charge $Z$ of the colloidal
particle. This new prefactor to the DLVO potential is known as the
renormalized
charge~\cite{Alexander,Trizac-renorm-PRL,Trizac-renorm-JCP}.

For spheroidal colloidal particles, we expect that, at large distances,
the potential behaves as~(\ref{eq:potential-anisotrop})
\begin{equation}
  \label{eq:F}
  \Psi(r,\theta,\phi)\sim F(\tildea,\theta,\phi) \,\frac{e^{-\kappa
      r}}{r}
  \,.
\end{equation}
The determination the anisotropy function $F$ requires to apply the
boundary condition at the surface of the colloid. But in this region
the linear solution is not valid. Therefore the anisotropy function
will be in principle different from the one obtained in the previous
section. For a spherical symmetric colloidal particle, the anisotropy
function is constant, and in the linear case, for impenetrable spheres
of radius $R$ it is $f_0=e^{\kappa R}/(1+\kappa R)$. Writing down the
anisotropy function $F$ (here constant) for the nonlinear situation as
\begin{equation}
  F = Z_{\text{ren}} l_B f_0 
  \,,
\end{equation}
defines a renormalized charge $Z_{\text{ren}}$ for the colloidal
particle, which encodes all the nonlinear phenomena due to its large
charge. In the anisotropic case, one might be tempted to write
\begin{equation}
  \label{eq:FZren}
  F(\tildea,\theta,\phi) = Z_{\text{ren}} l_B
  f_{\text{nl}}(\tildea,\theta,\phi)
  \,.
\end{equation}
However, there is no guaranty that the nonlinear effects would only
affect the anisotropy function by an overall multiplicative factor
(the renormalized charge), they would probably also affect the
dependence on the angular variables $(\theta, \phi)$,
i.e. $f_{\text{nl}}(\tildea,\theta,\phi)$ will be different from its
linear counterpart. Therefore the full anisotropy function $F$ is
``renormalized'' by the nonlinear effects, and the separation proposed
in~(\ref{eq:FZren}) is not unique.

For disc shaped colloidal particles, in~\cite{Agra}, the following
approach was used. Since the linear anisotropy function satisfies
$f(\tildea,0)=1$ in the direction perpendicular to the disc, the same
relation is proposed to define nonlinear anisotropy function
$f_{\text{nl}}(\tildea,0)=1$. From this, a unique definition of the
renormalized charge follows, using Eq.~(\ref{eq:FZren}) when
$\theta=0$. From a numerical resolution of the nonlinear
Poisson--Boltzmann equation one can obtain the anisotropy function for
the disc in the nonlinear regime.

For spheroids other that the disc or the rod, the property
$f(\tildea,\theta)=1$ does not hold for any direction
$\theta$. Therefore there is not an unambiguous way to define a
renormalized charge $Z_{\text{ren}}$ and a nonlinear anisotropic
function $f_{\text{nl}}$. Instead of trying to separate the nonlinear
effects into a renormalized charge and an anisotropy function, we will
study the combined effect in the complete anisotropy function $F$
defined by equation~(\ref{eq:F}).

To find the explicit form of the anisotropy function $F$, in
principle, one needs to solve the nonlinear equation and match the
linear solution~(\ref{eq:F}) valid only at large distances, with the
nonlinear one. This can always be done numerically. However an
analytical approach was proposed in~\cite{Trizac-renorm-PRL,
  Trizac-renorm-JCP, Agra}, valid for highly charged colloidal
particles with dimensions larger than the Debye length ($\kappa R\gg
1$). Under these conditions, the region where the nonlinear equation
should be used is very close to the colloidal particle, and at such
close distances, the colloidal particle can be seen as a planar object
(as a first approximation). Therefore one can approximate the
nonlinear potential in that region with the known planar solution of
the nonlinear Poisson--Boltzmann equation in one dimension. Then, the
linear potential valid at larger distances is matched to the nonlinear
one through an effective boundary condition of the Dirichlet type at
the surface of the colloidal particle,
i.e.~$\Psi(\xi_0,\eta,\phi)=\Psi_{0}$. The effective value $\Psi_0$ of
the potential is obtained from the solution of the nonlinear
one-dimensional Poisson--Boltzmann equation. Provided that the charge
of colloidal particle is large, $\Psi_0$ does not depend on the charge
of the colloidal particle, only on the constitution of the
electrolyte, i.e.~on the valencies $z_{+}$ and $z_{-}$. For a
charge-symmetric electrolyte, $z_{+}=z_{-}=1$, it is $\Psi_0=4$. For
further details on this picture of highly charged colloidal particles
seen as objects at constant potential, we refer the reader to
references~\cite{Trizac-renorm-PRL, Trizac-renorm-JCP, Agra}.

In section~\ref{subsec:constant-pot}, we solved the linear
Poisson--Boltzmann equation with the constant surface potential
boundary condition. From the analysis of that section, we obtain
directly an analytic expression for the renormalized anisotropy
function. For prolate spheroids,
\begin{equation}
  F(\tildea,\theta)_{pro}=\Psi_0 \sum_{l} \frac{2l+1}{2 \kappa
    R_{l}^{0}(i\tildea,\xi_0)} C_{l}(i\tildea)\,
    ps_{l}^{0}(i\tildea,\cos\theta)
    \,,
\end{equation}
and for oblate spheroids,
\begin{equation}
  F(\tildea,\theta)_{obl}=\Psi_0 \sum_{l} \frac{2l+1}{2 \kappa
    R_{l}^{0}(\tildea,i\xi_0)} C_{l}(\tildea)\,
    ps_{l}^{0}(\tildea,\cos\theta)
    \,.
\end{equation}

In the following section, we test the constant surface potential
boundary condition, by solving numerically the nonlinear
Poisson--Boltzmann equation for spheroids.

\subsection{Numerical resolution of Poisson--Boltzmann equation}

We consider a single spheroid with surface charge density $\sigma$
immersed in an electrolyte, and the spheroid is penetrable by the
electrolyte. In the mean field approximation, the screened
electrostatic potential satisfies the nonlinear Poisson--Boltzmann
equation~(\ref{PBnl}) outside and inside the spheroid. This
differential equation is complemented by the boundary conditions
$\nabla \Psi(\xi,\eta,\phi) \to 0 $ when $\xi\to\infty$, $\Psi$
continuous at the surface of the spheroid $\xi=\xi_0$, and the
discontinuity of the electric field at the surface of the spheroid is
proportional to $\sigma$, Eq.~(\ref{eq:discont-psi-prime}). To solve
equation~(\ref{PBnl}) numerically, we implemented an algorithm similar
to the one used in Refs.~\cite{Chapot-Bocquet-Trizac2, Agra} for rods
and discs.

Let $\rho_{\text{col}}(\xi)= \sigma \delta(\xi-\xi_0)/h_{\xi}$ be the
charge density of the colloidal particle, with $h_{\xi}$ the scale
factor corresponding to the $\xi$ coordinate~\cite{Morse},
$h_{\xi}=a\sqrt{(\xi^2-\eta^2)/(\xi^2-1)}$ for prolate spheroids,
$h_{\xi}=a\sqrt{(\xi^2+\eta^2)/(\xi^2+1)}$ for oblate spheroids.
Poisson--Boltzmann equation~(\ref{PBnl}), with the boundary conditions
imposed on $\Psi$, can be rewritten as
\begin{equation}
  \label{PBnl2}
  \nabla^2 \Psi - \kappa^{2} \Psi =  \frac{\kappa^2}{z_{+}+z_{-}}
  \left[ e^{z_{-} \Psi}- e^{-z_{+} \Psi} \right] - \kappa^{2} \Psi
  -4\pi (l_B/e) \rho_{\text{col}}
  \,.
\end{equation}
Introducing the kernel $G$ of $\nabla^2 -\kappa^2$ in free space,
\begin{equation}
  G(\r,\r')=-\frac{1}{4\pi} \frac{e^{-\kappa |\r-\r'|}}{|\r-\r'|}
  \,,
\end{equation}
equation~(\ref{PBnl2}) can be cast as an integral equation
\begin{equation}
  \label{eq:PBint}
  \Psi(\r)=\int_{\mathbb{R}^3} G(\r,\r') \left[
    \frac{\kappa^2}{z_{+}+z_{-}} \left[ e^{z_{-} \Psi(\r')}- e^{-z_{+}
        \Psi(\r')} \right] - \kappa^{2} \Psi(\r') -4\pi (l_B/e)
    \rho_{\text{col}}(\r')\right] \,d\r'
\end{equation}
The numerical algorithm to solve this equation is iterative. It starts
with a trial value for the potential, $\Psi_{\text{in}}$, computes the
integral in the right-hand side of equation~(\ref{eq:PBint}) to obtain
a new value for the potential, $\Psi_{\text{out}}$, then uses a linear
mixing of the two, $(1-\alpha)\Psi_{\text{in}} +
\alpha\Psi_{\text{out}}$, to start the next iteration, with the
parameter $\alpha$ in the range $10^{-4}$ -- $10^{-1}$. This procedure
is iterated until convergence is achieved, i.e., $\int
|\Psi_{\text{in}}-\Psi_{\text{out}}| / \int |\Psi_{\text{in}}| <
\epsilon$, with $\epsilon$ a relative error tolerance of the order
$10^{-5}$.

For the present problem, we used spheroidal coordinates. The kernel
$G$ can be expanded in spheroidal wave functions as
\begin{equation}
  G(\xi,\eta,\phi;\xi',\eta',\phi')=
  -\frac{\kappa}{4\pi}
  \sum_{l,m} (2l+1) e^{im(\phi-\phi')}
  ps_{l}^{m} (i\tildea,\eta)  ps_{l}^{m} (i\tildea,\eta')
  i^{-l} S_{l}^{m(1)}(i\tildea,\xi_{<}) R_{l}^{m}(i\tildea,\xi_{>})
\end{equation}
for a prolate spheroid, and
\begin{equation}
  G(\xi,\eta,\phi;\xi',\eta',\phi')=
  -\frac{\kappa}{4\pi}
  \sum_{l,m} (2l+1) e^{im(\phi-\phi')}
  ps_{l}^{m} (\tildea,\eta)  ps_{l}^{m} (\tildea,\eta')
  i^{-l} S_{l}^{m(1)}(\tildea,i\xi_{<}) R_{l}^{m}(\tildea,i\xi_{>})
\end{equation}
for an oblate spheroid, with $\xi_{<}=\min(\xi,\xi')$ and
$\xi_{>}=\max(\xi,\xi')$. Note that, since the charge density of the
colloidal particle does not depend on the azimuthal angle, in the
integral equation~(\ref{eq:PBint}), only the terms for $m=0$ remain
after the integral over the azimuthal angle $\phi'$ is performed.

Numerically, one cannot compute the integral in
equation~(\ref{eq:PBint}) in the whole space $\mathbb{R}^{3}$, but
rather it has to be restricted to a finite region
$\xi<\xi_{\max}$. The integral equation~(\ref{eq:PBint}) with the
integral restricted to a finite region $\xi<\xi_{\max}$, is equivalent
to the nonlinear Poisson--Boltzmann differential
equation~(\ref{PBnl2}), for $\xi<\xi_{\max}$, with the boundary
condition that outside the region of integration ($\xi>\xi_{\max}$)
the potential satisfies the linear Poisson--Boltzmann
equation~(\ref{helmh}). This is perfectly acceptable provided that
$\tildea\xi_{\max}\gg1$, since due to the screening we know that the
potential will be small, $\Psi\ll1$, at large distances from the
colloidal particle, and thus it will satisfy the linear
Poisson--Boltzmann equation.

\subsection{Numerical results}

\begin{figure}[ht]
\centering
\vspace{8mm}
\includegraphics[width=\GraphicsWidth]{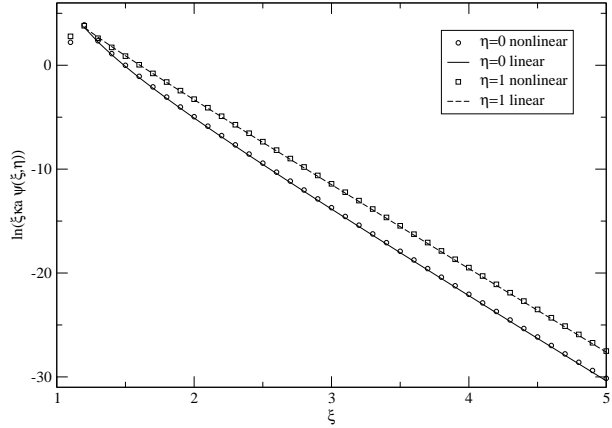}
\vspace{3mm}
\caption{Electrostatic potential for a prolate spheroid with
  $\xi_0=1.2$, $\tildea=8.0$, and surface charge density given by
  $l_B\sigma/(\kappa e) = 10$. The symbols are from a numerical
  resolution of the nonlinear Poisson--Boltzmann equation, and the
  lines from the solution of the linear Poisson--Boltzmann equation
  with the fixed surface potential boundary condition.}
\label{fig:nl-pot-prol}
\end{figure}

Figure~\ref{fig:nl-pot-prol} shows the numeric solution of the
nonlinear Poisson--Boltzmann equation for a prolate spheroid with
$\xi_0=1.2$ and $\tildea=8.0$, $\kappa R= 9.6$, $\kappa R'=5.3$, and
aspect ratio $R'/R=0.55$. Its surface charge density is given by
$l_B\sigma/(\kappa e) = 10$. The corresponding bare charge is then
given by $Z l_B/a \simeq 690 \gg 1$, which is large, thus in the
nonlinear and saturated regime. The spheroid is immersed in a
charge-symmetric electrolyte with $z_{+}=z_{-}=1$. In the same figure,
we show the solution of the linear Poisson--Boltzmann equation, with
the effective boundary condition of fixed potential $\Psi_0=4$ at the
surface of the spheroid
\begin{equation}
  \Psi_{\text{lin}}(\xi,\eta)= 
  \Psi_0\sum_{l} \frac{2l+1}{2R_l^0(i\tilde{a},\xi_0)}\,C_{l}(i\tildea)
  \,
  ps_{l}^{0}(i\tildea,\eta) R_{l}^{0}(i\tildea,\xi)
  \,.
\end{equation}
At larges distances from the spheroid, $\kappa a \xi \geq 2$, the
linear approximation is excellent. This strongly supports the picture
of constant potential objects for highly charged colloidal particles
presented in the previous section and in
Refs.~\cite{Trizac-renorm-PRL, Trizac-renorm-JCP, Agra}.

\begin{figure}[ht]
\centering
\vspace{10mm}
\includegraphics[width=\GraphicsWidth]{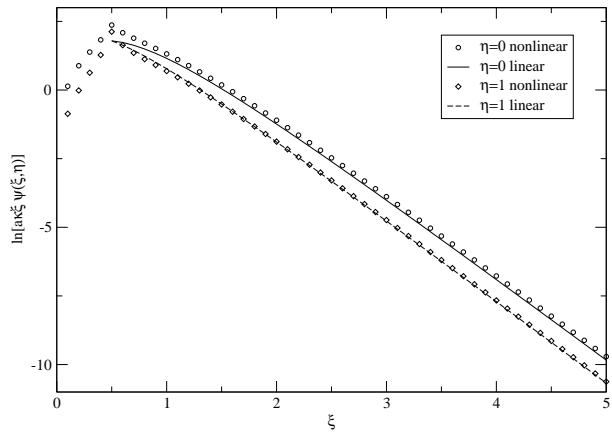}
\vspace{3mm}
\caption{Electrostatic potential for an oblate spheroid with
  $\xi_0=0.50$, $\tildea=3.0$, and surface charge density given by
  $l_B\sigma/(\kappa e) = 10$. The electrolyte is charge-symmetric
  with valencies $z_{+}=1$ and $z_{-}=1$. The symbols are from a
  numerical resolution of the nonlinear Poisson--Boltzmann equation, and
  the lines from the solution of the linear Poisson--Boltzmann
  equation with the fixed surface potential boundary condition.}
\label{fig:nl-pot-obl-sym}
\end{figure}

\begin{figure}[ht]
\centering
\vspace{3mm}
\includegraphics[width=\GraphicsWidth]{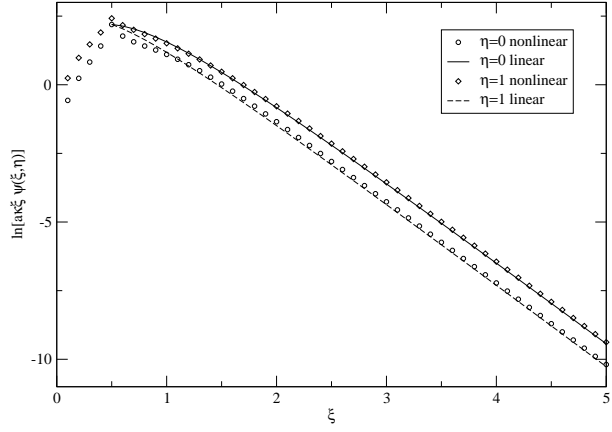}
\vspace{3mm}
\caption{Electrostatic potential for an oblate spheroid with
  $\xi_0=0.50$, $\tildea=3.0$, and surface charge density given by
  $l_B\sigma/(\kappa e) = 10$. The electrolyte has valencies $z_{+}=2$
  and $z_{-}=1$. The symbols are from a numerical resolution of the
  nonlinear Poisson--Boltzmann equation, and the lines from the
  solution of the linear Poisson--Boltzmann equation with the fixed
  surface potential boundary condition.}
\label{fig:nl-pot-obl-21}
\end{figure}

\begin{figure}[ht]
\centering
\vspace{3mm}
\includegraphics[width=\GraphicsWidth]{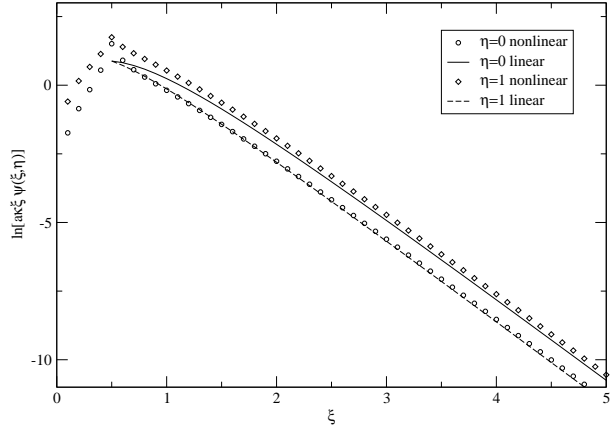}
\vspace{3mm}
\caption{Electrostatic potential for an oblate spheroid with
  $\xi_0=0.50$, $\tildea=3.0$, and surface charge density given by
  $l_B\sigma/(\kappa e) = 10$. The electrolyte has valencies $z_{+}=1$
  and $z_{-}=2$. The symbols are from a numerical resolution of the
  nonlinear Poisson--Boltzmann equation, and the lines from the
  solution of the linear Poisson--Boltzmann equation with the fixed
  surface potential boundary condition.}
\label{fig:nl-pot-obl-12}
\end{figure}

In figure~\ref{fig:nl-pot-obl-sym}, we present the results for an
oblate spheroid, now with $\xi_0=0.50$, $a=3.0$, $\kappa R'=1.5$,
$\kappa R=3.4$, $R'/R=0.45$ and charge density $l_B\sigma/(\kappa e) =
10$. The agreement with the linear solution for fixed surface
potential $\Psi_0$ is again very good. Since the dimensions of the
spheroid are not very large compared to the Debye length, one could
eventually search for size dependent (in $\kappa R$ and $\kappa R'$)
corrections to the planar value $\Psi_0=4$ to improve the linear
approximation. As an additional test, in
figure~\ref{fig:nl-pot-obl-21} and~\ref{fig:nl-pot-obl-12}, we present
the results for a charge-asymmetric electrolytes with $z_{+}=2$,
$z_{-}=1$, (2:1), and the reverse situation $z_{+}=1$, $z_{-}=2$,
(1:2). In these cases, the saturation values of the effective
potential are $\Psi_0=6$ for the (2:1) situation, and
$\Psi_0=6(2-\sqrt{3})$ for the (1:2)
situation~\cite{Tellez-Trizac-asym}. Again, the agreement between the
nonlinear solution and the linear one far from the colloidal particle
is excellent, in all directions (in the figures we report only the
results for the directions $\eta=0$, i.e. $\theta=\pi/2$ and $\eta=1$,
i.e. $\theta=0$).

\section{Monte Carlo simulations}\label{Sim}
In an attempt to study the charged spheroidal colloid system beyond
mean field theories, we made some preliminary simulations of a system
of hard spheroidal colloids with a point charge $Ze$ at their centers,
and explicit coions and counter-ions of charge $+e$ and $-e$ confined in
a simulation box of side length $L$, and using periodic boundary
conditions. The reason to choose the model with a charge at the center
of the particles rather than with constant potential or charge density
at their surface was a matter of computational efficiency, as in the
latter cases the analytic expression for the interaction between the
particles is very complicated (except for the spherical case) and thus
the potential needs to be computed numerically. On top of that, as we
use periodic boundary conditions, it would be necessary to calculate
the Ewald sums form of the potential to take correctly into account
the images of the system, which makes it even more difficult, and
numerically time consuming.

Although the simulations were carried out at relatively low colloid
densities, with a colloid volume fraction of 0.0025, the systems are
not at the infinite dilution limit, and five colloids were simulated
in the simulation box, rather than one, to avoid the bias that a
crystalline structure could introduce to the system.  Unfortunately,
the analytical solution of Poisson-Boltzmann equation in the case of
point charges at the center of the spheroidal colloids turns out to be
difficult even in its linearized (Debye-H\"uckel) form, as the
symmetry of the potential inside and outside of the spheroidal
particle is not the same. For these reasons, we will not attempt a
direct quantitative comparison of the simulation results with the
solution of Poisson-Boltzmann equation, but rather to make a
qualitative analysis.

\subsection{Model}
The particles interact at short range with the spheroidal hard core
potential, and at long range with the Coulomb electrostatic potential.
The dimensionless long range Coulomb interaction potential ($\beta
e\times$ potential) between particles $i$ and $j$ is given by
\begin{equation}
 \mathcal{U}_C(r_{ij})=\frac{z_iz_jl_B}{r_{ij}},
\end{equation}
where $z_i$ and $z_j$ are the valences of particles $i$ and $j$ and
$r_{ij}=|\mathbf{r}_j-\mathbf{r}_i|$ is the distance between the
centers of the particles. As we use periodic boundary conditions, in
order to take correctly into account the interactions between
particles in the main simulation box and particles in the periodic
images, Ewald sums were used, so the Coulomb potential energy takes
the form \cite{Allen}
\begin{eqnarray}
{E}_{C}&=&\frac{1}{2}\sum_{ij}\left[\sum_{\mathbf{n}}'z_iz_jl_B\frac{\mbox{erfc}(\alpha
    \tilde{r}_{ij})}{\tilde{r}_{ij}}-\frac{1}{L^3}\sum_{\mathbf{k}}z_iz_jl_B\left(\frac{4\pi}{k^2}\right)e^{-\frac{k^2}{4\alpha^2}}\ e^{i(\mathbf{k}\cdot\mathbf{r}_{ij})}\right]-\frac{\alpha}{\sqrt{\pi}}\sum_iz_i^2l_B,
\label{Ewcoul}
\end{eqnarray}
where $\mathbf{n}=(n_x,n_y,n_z)\in\mathbb{Z}^3$ is the image vector in
real space, $\tilde{r}_{ij}=|\mathbf{r}_{ij}+L\mathbf{n}|$, the prime
denotes that $i\neq j$ if $\mathbf{n}=\mathbf{0}$,
$\mathbf{k}=2\pi\mathbf{p}/L$, with $\mathbf{p}\in\mathbb{Z}^3$, is
the grid vector in the Fourier space and $\alpha$ is a parameter
controlling the convergence of the real and Fourier space sums. We use
the minimum image convention so that $\mathbf{n}=\mathbf{0}$. The
Fourier sum is truncated at $|\mathbf{p}|=9$ with $\alpha= 0.76$.\\

The short ranged (hard-core) potential is
\begin{equation}
\mathcal{U}_{HC}(\mathbf{r}_{ij},\Omega_{i},\Omega_{j})=\left\{\begin{array}{ll}
0;&F(\mathbf{r}_{ij},\Omega_i,\Omega_j)\geq1\\
\infty;&F(\mathbf{r}_{ij},\Omega_i,\Omega_j)<1
\end{array}\right.
\end{equation}
where $\mathbf{r}_{ij}$ is
the separation vector between the centers of the ellipsoids and $F(\mathbf{r}_{ij},\Omega_i,\Omega_j)$ is the contact function between two
ellipsoids with orientations $\Omega_i$ and $\Omega_j$ as defined in
\cite{Perram}. This function $F$ has the property that for two ellipsoids A
and B
\begin{equation}
F(\mathbf{r}_{AB},\Omega_A,\Omega_B)=\left\{\begin{array}{ll}
<1& \mbox{if A and B overlap}\\
=1& \mbox{if A and B are externally tangent}\\
>1& \mbox{if A and B do not overlap}\\
\end{array}\right.
\end{equation}

\begin{figure}[ht]
\centering
\includegraphics[width=\GraphicsWidth,angle=-90]{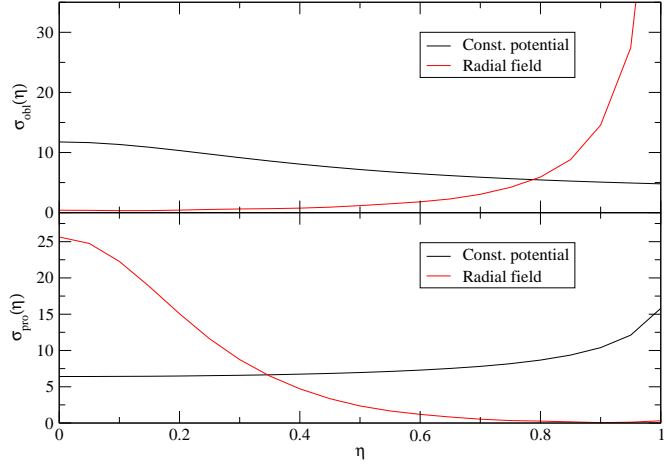}
\caption{Surface charge density as a function of $\eta$ for the
  oblate ($R'/R=1/3$, upper)
  and prolate ($R'/R=1/3$, lower) cases. The constant potential case was calculated with
  $\Psi_0=4$ and the radial field case with $Ze=30$.}
\label{fig_Cdens}
\end{figure}

Each spheroidal particle has a point charge at its
center. Alternatively, one can think that each spheroid is void inside
and has a certain charge density in its surface that creates an
electrostatic potential outside the spheroid, at a distance $r$ from
its center, proportional to $1/r$. Now, we compute this equivalent
surface charge, which could be used in principle as a Neumann boundary
condition to solve analytically the Poisson-Boltzmann equation. The
Green function of Laplace equation, in spheroidal coordinates
\cite{Morse}, is 
\begin{eqnarray}
\frac{1}{|\mathbf{r}-\mathbf{r}'|}=\frac{1}{a}\sum_{n=0}^\infty(2n+1)\sum_{m=0}^n\epsilon_mi^m\left[\frac{(n-m)!}{(n+m)!}\right]^2\cos[m(\phi-\phi')]\nonumber\\
\times P_n^m(\eta')P_n^m(\eta)\left\{\begin{array}{ll}
P_n^m(\xi')Q_n^m(\xi);&\xi>\xi'\\
P_n^m(\xi)Q_n^m(\xi');&\xi<\xi'
\end{array}\right.
\end{eqnarray}
in the prolate case, and
\begin{eqnarray}
\frac{1}{|\mathbf{r}-\mathbf{r}'|}=\frac{1}{a}\sum_{n=0}^\infty(2n+1)\sum_{m=0}^n\epsilon_mi^{m+1}\left[\frac{(n-m)!}{(n+m)!}\right]^2\cos[m(\phi-\phi')]\nonumber\\
\times P_n^m(\eta')P_n^m(\eta)\left\{\begin{array}{ll}
P_n^m(i\xi')Q_n^m(i\xi);&\xi>\xi'\\
P_n^m(i\xi)Q_n^m(i\xi');&\xi<\xi'
\end{array}\right.
\end{eqnarray}
in the oblate case, with $P_{n}$ and $Q_{n}$ the Legendre functions of
degree $n$ of the first and second kind respectively. The value of
$\epsilon_m$ is $1$ if $m=0$, and $2$ otherwise. For the case of
spheroids of revolution, and taking $\mathbf{r}'$ at the origin, we have
that only the $m=0$ terms are non zero. Then, the potential inside and
outside of a spheroidal particle is
\begin{eqnarray}
\mathcal{U}_{out}(\xi,\eta)=\frac{Z l_B}{r}&=&\frac{Zl_B}{a}\sum_{n=0}^\infty(2n+1)P_n(0)P_n(\eta)Q_n(\xi),\label{outP}\\
\mathcal{U}_{in}(\xi,\eta)&=&\sum_{n=0}^\infty A_nP_n(\eta)P_n(\xi).\label{inP}
\end{eqnarray}
in the prolate case, and
\begin{eqnarray}
\mathcal{U}_{out}(\xi,\eta)=\frac{Z
  l_B}{r}&=&i\frac{Zl_B}{a}\sum_{n=0}^\infty(2n+1)P_n(0)P_n(\eta)Q_n(i\xi),\label{outO}\\ 
\mathcal{U}_{in}&=&\sum_{n=0}^\infty
A_nP_n(\eta)P_n(i\xi).\label{inO}
\end{eqnarray}
in the oblate case. The constants $A_{n}$ are determined by using the
boundary condition
\begin{equation}
\mathcal{U}_{in}(\xi_0)=\mathcal{U}_{out}(\xi_0)
\,.
\end{equation}
Then, using
\begin{equation}
\frac{1}{h_\xi}\left.\Dd{\mathcal{U}_{out}}{\xi}\right|_{\xi=\xi_0}-\frac{1}{h_\xi}\left.\Dd{\mathcal{U}_{in}}{\xi}\right|_{\xi=\xi_0}=-4\pi
l_B\frac{\sigma(\eta)}{e}
\end{equation}
along with equations (\ref{outP}) to (\ref{inO}), it is possible to find a surface charge
density equivalent to having a point charge $Ze$ at the center of the
spheroids (i.e.~the charge density that produces a potential
outside the spheroidal particle that is proportional to $1/r$)
\begin{eqnarray}
\sigma(\eta)=\frac{Ze}{4\pi a^2}\frac{1}{\sqrt{(\xi_0^2-\eta^2)(\xi_0^2-1)}}\sum_{n=0}^\infty(4n+1)(-1)^n\frac{(2n-1)!!}{(2n)!!}\frac{P_{2n}(\eta)}{P_{2n}(\xi_0)}.
\end{eqnarray}
for the prolate case, and
\begin{eqnarray}
\sigma(\eta)=\frac{Ze}{4\pi a^2}\frac{1}{\sqrt{(\xi_0^2+\eta^2)(\xi_0^2+1)}}\sum_{n=0}^\infty(4n+1)(-1)^n\frac{(2n-1)!!}{(2n)!!}\frac{P_{2n}(\eta)}{P_{2n}(i\xi_0)}.
\end{eqnarray}
for the oblate case. Here the difficulty of solving analytically the
linear PB equation (eq. (\ref{helmh})) becomes evident, as the surface
charge density is given in terms of the Legendre functions, while
solution of the Helmholtz equation in spheroidal coordinates are the
spheroidal wave functions.  It is worth mentioning here that this
charge distribution, which produces a radial field outside of the
particle and the charge distribution associated with a constant
surface potential (eq. (\ref{cpCdens})) have different behaviors
(Figure \ref{fig_Cdens}): the radial field charge distribution is
higher in the direction of the small axis, while constant potential
surface charge is higher in the direction of the large axis.

\subsection{Simulations}
Monte Carlo simulations were carried out for prolate and oblate colloids with
$R'/R=1/3$, and three different values of $Z=$ $-50$, $-200$ and $-400$ for each
case. The Bjerrum length used is $l_B^*=0.0625$ (in this section the lengths are normalized as
$r^*=r/(8RR'^2)^{1/3}$ in the prolate case and $r^*=r/(8R'R^2)^{1/3}$ in the oblate case). Each system
consisted of 5 spheroidal colloidal particles, with a packing fraction
$\eta=0.0025$, 500 coions and 750, 1500 and 2500 counterions respectively. The
reduced volume is $V^*=1028.19$.

From the simulations the colloid-counterion correlation functions $h=g-1$
\begin{equation}
h(r_{ij},\Omega_i,\Omega_j)=\sum_{m}\sum_{n}\sum_{l}h^{mnl}(r_{ij})\hat{\Phi}_{00}^{mnl}(\Omega_i,\Omega_j,\hat{\mathbf{R}}),
\label{coefll}
\end{equation}
were obtained. The interparticle distance is $r_{ij}$, $\Omega_i$ are
the Euler angles that give the orientation of particle $i$,
$\hat{\mathbf{R}}$ is the orientation of the interparticle vector, and
the $\hat{\Phi}_{\mu'\nu'}^{mnl}$ are the rotational invariants
defined in \cite{Blum} multiplied by a prefactor
\begin{equation}
f^{mnl}=l!/\left(\begin{array}{lll}
m&n&l\\
0&0&0
\end{array}\right),
\end{equation}
where the term on the denominator is a Wigner's 3j symbol.

In our case, since the microions have spherical symmetry and the
spheroids have rotational symmetry around their revolution axis as
well as the perpendicular plane to this axis that crosses at their
center, the expression (\ref{coefll}) reduces to a simple expansion in
Legendre polynomials
\begin{equation}
h(r,\theta)=\sum_{l=0}^{\infty}h^{2l}(r)P_{2l}(\cos\theta),
\label{correl}
\end{equation}
where $\theta$ is the angle of the interparticle vector with respect to the
orientation of the revolution axis of the colloid and  
\begin{equation}
h^{2l}(r)=\frac{4l+1}{2}\int_{-1}^1P_{2l}(\cos\theta)h(r,\theta)d(\cos\theta).
\end{equation}
The radial functions
$h^{2l}(r)$ were obtained from the simulations and then used to reconstruct the
correlation function using equation (\ref{correl}), truncating the summation at $l_{max}=100$.

The potential of mean force, $\mathcal{U}_{mf}$, associated with the
colloid plus its screening cloud of ions was obtained from the
correlation functions using
\begin{equation}
h(r,\theta)=e^{-\mathcal{U}_{mf}(r,\theta)}-1.
\end{equation}
The simulations with 1250--3000 ions were equilibrated for
100000--150000 Monte Carlo steps and averaged for 60000--130000 Monte
Carlo steps.

\begin{figure}[ht]
\centering
\includegraphics[width=\GraphicsWidth]{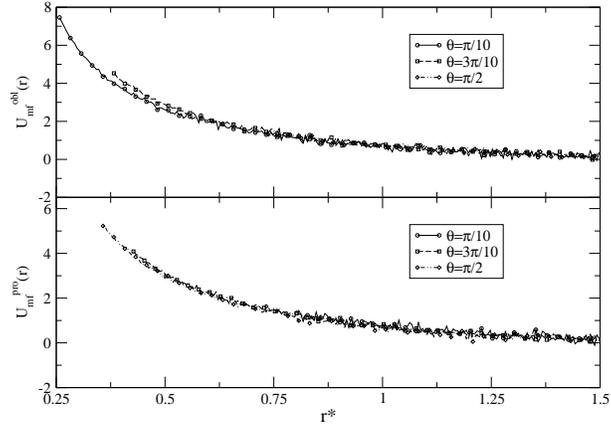}
\vspace{3mm}
\caption{Effective potential vs. $r$ for the oblate (up) and
  prolate (down) cases with $R'/R=1/3$ for different angles with $Z=50$.}
\label{fig_pot1}
\
\vspace{3mm}
\end{figure}

\begin{figure}[ht]
\centering
\includegraphics[width=\GraphicsWidth]{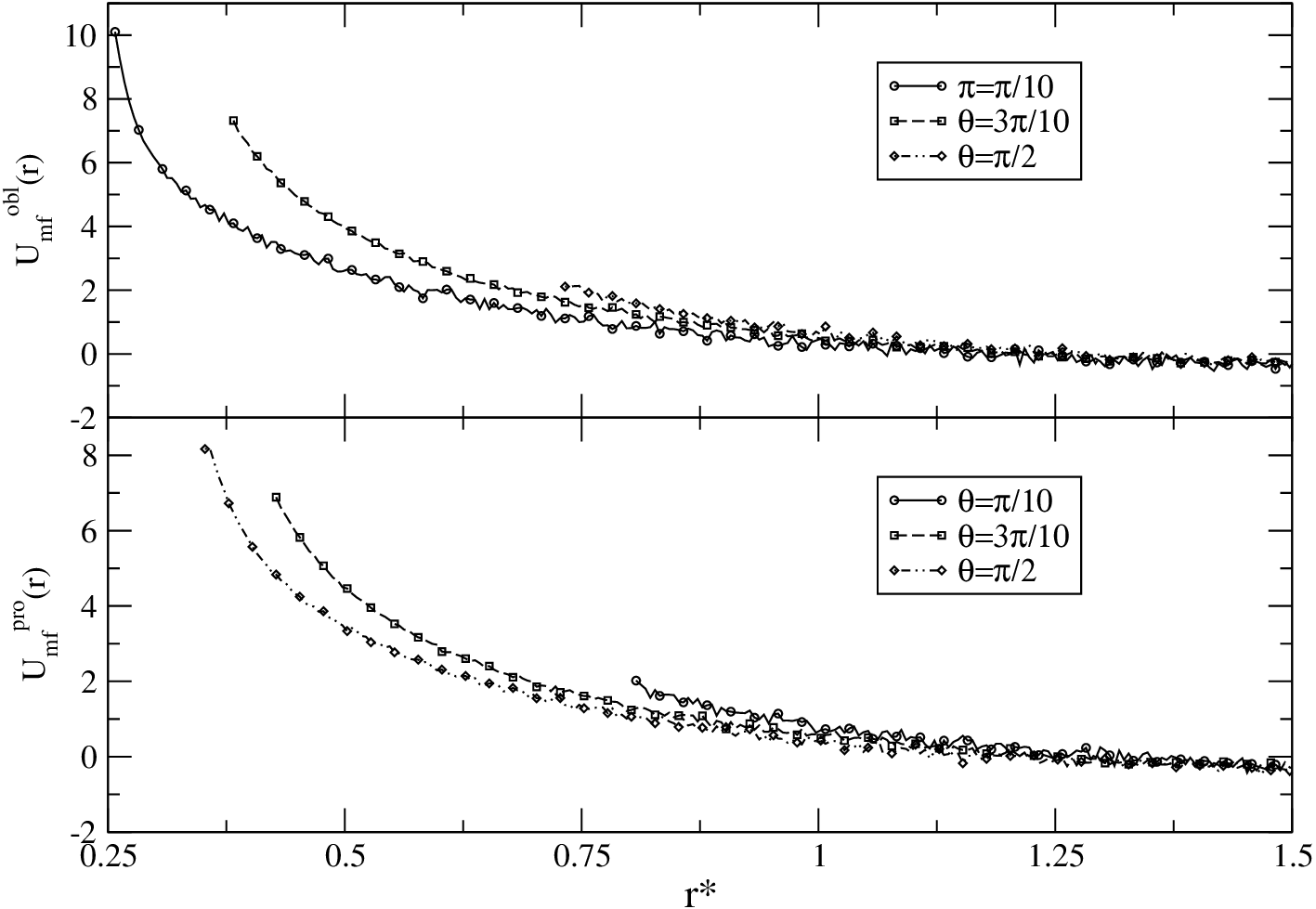}
\
\vspace{3mm}
\caption{Effective potential vs. $r$ for the oblate (up) and
  prolate (down) cases with $R'/R=1/3$ for different angles with $Z=200$.}
\label{fig_pot2}
\end{figure}

\begin{figure}[ht]
\centering
\
\vspace{3mm}
\includegraphics[width=\GraphicsWidth]{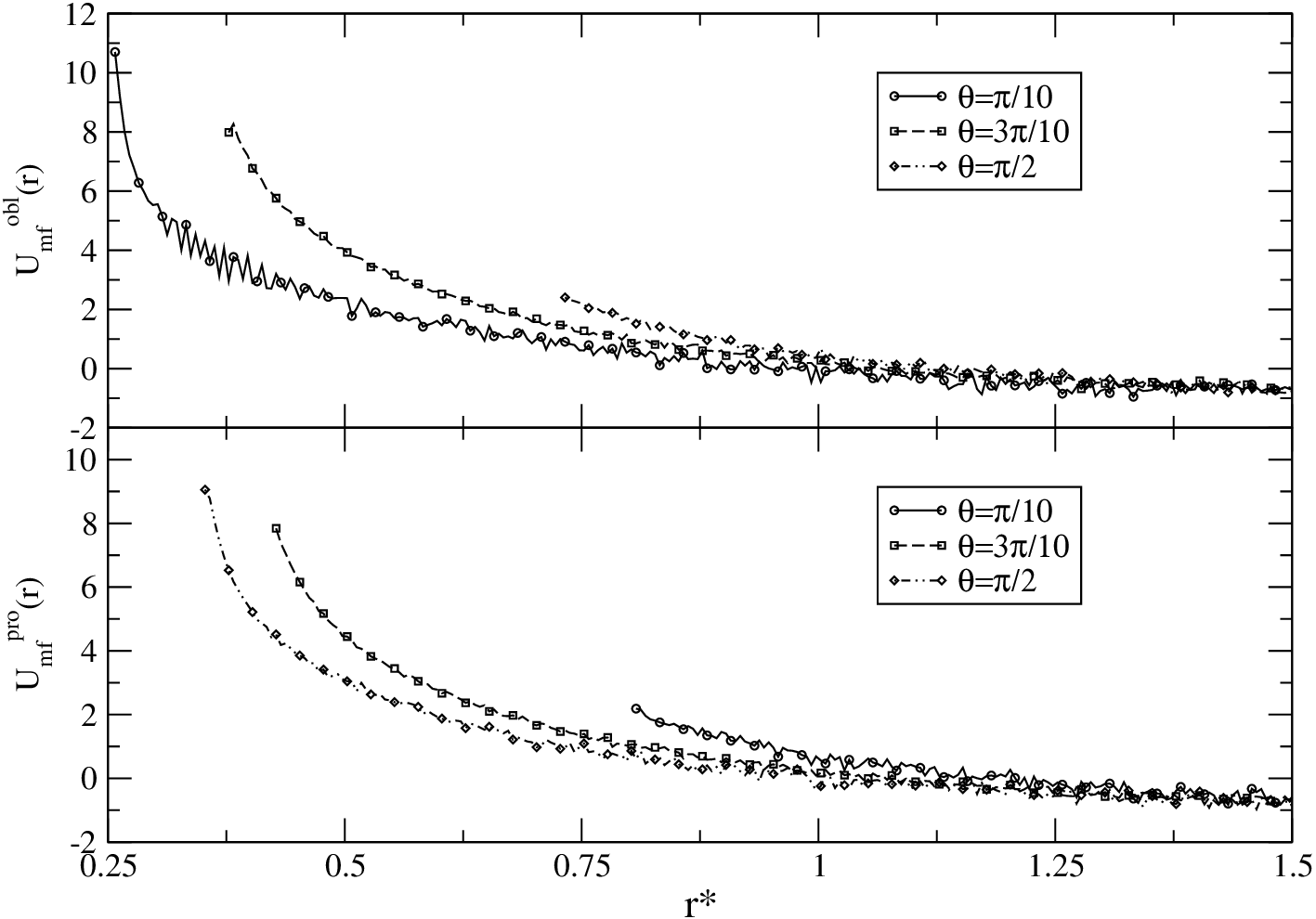}
\caption{Effective potential vs. $r$ for the oblate (up) and
  prolate (down) cases with $R'/R=1/3$ for different angles with $Z=400$.}
\label{fig_pot3}
\end{figure}

The potential of mean force is plotted against $r^*$ in figures \ref{fig_pot1},
\ref{fig_pot2} and \ref{fig_pot3}. In the figures it can be seen that for the
oblate case the potential increases as we go from the poles of the colloidal
particles ($\theta=0$) to the equatorial line ($\theta=\pi/2$) for a fixed
value of $r^*$. For the weak
coupling case ($Z=50$) the potential depends little on the orientation, as
could be expected for the isotropic (spherical) case, but the effect is more evident as $Z$ increases. This is the same
behavior that was observed for the solution of the linearized Poisson-Boltzmann
in the constant charge and constant potential cases (figures \ref{anisQ} and \ref{anis}), even though the
distribution of charge is very different in all cases (figure \ref{fig_Cdens}). 

In the prolate case the potential decreases as we go from $\theta=0$
to $\theta=\pi/2$ for fixed $r^*$, and it is again the same behavior
observed for the solution of the linearized Poisson-Boltzmann in the
constant charge and constant potential cases.

From these results it can be seen that the short range steric
interaction (the shape of the excluded volume) has a strong effect on the
effective screened potential of the colloid, producing an anisotropy which is
larger in the orientation where the curvature of the colloid is higher, even
for different surface charge distributions.

\section{Conclusion}

We studied the effective electrostatic potential created by a
spheroidal colloidal particle and its screening cloud, within the mean
field approximation, using Poisson Botzmann equation in its linear and
nonlinear forms. Also, we did a preliminary exploration beyond the
mean field with Monte Carlo simulations. 

In all the studied cases, we confirmed the persistence of anisotropy
of the effective potential at large distances, as it was already
noticed for other anisotropic particles such as cylinders and
discs~\cite{Agra, Chapot-Bocquet-Trizac}. For highly charged colloidal
particles, but within the validity domain of mean field, we were
able to test the picture of constant potential objects for these
anisotropic particles~\cite{Trizac-renorm-PRL, Trizac-renorm-JCP}: at
large distances from the charged particle, the nonlinear solution of
Poisson--Boltzmann equation can be approximated by the linear one with
an effective boundary condition of constant potential at the surface
of the particle. In this anisotropic situation, our work has provided
a strong test for this picture, since a constant potential boundary
condition is very different from a constant surface charge density,
contrary to the situation in the spherical case.

With this constant potential boundary condition, we found that, at
large distances from the spheroidal particle, the potential is larger
in the direction where the curvature of the particle is higher, that
is the large axis direction. From the simulations, we observed that
the potential of mean force around the spheroidal colloidal particles
with a point charge a their centers, has a similar behavior to that
of the constant surface charge density and constant surface potential
cases computed analytically, even though the equivalent surface charge
density is very different from the two previous cases.

\begin{acknowledgements}
  The authors thank E.~Trizac for useful discussions. Partial
  finantial support from ECOS-Nord/COLCIENCIAS-MEN-ICETEX, and from
  Comit\'e de Investigaciones y Posgrados, Facultad de Ciencias,
  Universidad de los Andes is acknowledged.  The numerical
  computations were performed using IDRIS (Institut du D\'eveloppement
  et des Ressources en Informatique Scientifique) ressources under
  project no.~CP9 -- 092104.
\end{acknowledgements}

\bibliographystyle{unsrt}
\bibliography{paper} 

\end{document}